\documentclass{amsart}
\usepackage{amsfonts}
\usepackage{amsmath}
\usepackage{amssymb}
\usepackage{graphicx}

\theoremstyle{plain}
\newtheorem{theorem}{Theorem}[section]

\theoremstyle{definition}

\theoremstyle{remark}
\newtheorem{remark}[theorem]{Remark}
\numberwithin{equation}{section}

\input{tcilatex}

\begin{document}
\title[The AKNS hierarchy revisited]{The AKNS hierarchy revisited: A vertex
operator approach and its Lie-algebraic structure}
\author{Denis Blackmore$^{1}$}
\address{$^{1}$Department of Mathematical Sciences and Center for Applied
Mathematics and Statistics, New Jersey Institute of Technology, Newark, NJ
07102, USA}
\email{deblac@m.njit.edu}
\author{Anatoliy K. Prykarpatsky$^{2,3}$}
\address{$^{2}$The Department of Mining Geodesy and Environment Engineering
at the AGH University of Science and Technology, Krakow 30059, Poland\\
and\\
$^{3}$Department of Economical Cybernetics at the Ivan Franko Pedagogical
State University, Drohobych, Lviv region, Ukraine \ }
\email{pryk.anat@ua.fm, prykanat@cybergal.com}
\subjclass{Primary 58A30, 56B05 Secondary 34B15 }
\keywords{AKNS hierarchy, Lax integrability, Lie-algebraic approach, vertex
operators}
\date{present}

\begin{abstract}
A novel approach - based upon vertex operator representation - is devised to
study the AKNS hierarchy. It is shown that this method reveals the
remarkable properties of the AKNS hierarchy in relatively simple, rather
natural and particularly effective ways. In addition, the connection of this
vertex operator based approach with Lie-algebraic integrability schemes is
analyzed and its relationship with $\tau$-function representations is
briefly discussed.
\end{abstract}

\maketitle

\section{Introduction}

The \textquotedblleft miraculous\textquotedblright\ properties of the AKNS
hierarchy related to calculations connected with the integrability of
nonlinear dynamical systems have, since the early work of their discoverers 
\cite{AKNS,Ne}, been the focus of considerable research. These
investigations, such as in \cite{Di,FT,GW,No,PM,RS-T,BK1,BK2,MJD}, have
produced further insights into the nature of the AKNS hierarchy and several
additional methods of construction. In what follows, we devise an
alternative approach to exploring the properties of the AKNS hierarchy based
upon its generating vector field form and related vertex operator
representation. It appears that our formulation offers several advantages
over existing methods when it comes to simplicity, effectiveness,
flexibility and ease of extension, but \ more detailed confirmation of these
observations must await further investigations.

To set the stage for our approach, we begin with some fundamentals of the
remarkable sequence of Lax integrable dynamical systems that is the focus of
this study. We shall analyze the AKNS hierarchy of Lax integrable dynamical
systems on a complex $2\pi$-periodic functional manifold $M\subset C^{\infty
}(\mathbb{R}/2\pi\mathbb{Z};\mathbb{C}^{2})$, which is well known \cite%
{AKNS,FT,Ne,No} to be related to the following linear differential spectral
problem of Lax type:

\begin{equation}
df/dx-\ell (x;\lambda )f=0,\ \ \ell (x;\lambda ):=\left( 
\begin{array}{cc}
\lambda /2 & u \\ 
v & -\lambda /2%
\end{array}%
\right) .  \label{V1.1}
\end{equation}%
Here $x\in \mathbb{R},f\in L^{1}(\mathbb{R};\mathbb{C}^{2})$, the vector
function $(u,v)^{\intercal }\subset M$, $"^{\intercal }"$ denotes the
transpose and $\lambda \in \mathbb{C}$ is a spectral parameter. Assume that
a vector function $(u,v)^{\intercal }\subset M$ depends parametrically on
the infinite set $t:=\{t_{1},t_{2},t_{3},\ldots \}\in \mathbb{C}^{\mathbb{N}%
} $ in such a way that the generalized Floquet spectrum $\sigma (\ell
):=\{\lambda \in \mathbb{C}:\sup_{x\in \mathbb{R}}||f(x;\lambda
)||_{1}<\infty \}$ of the problem (\ref{V1.1}) persists in being
parametrically iso-spectral, that is $d\sigma (\ell )/dt=0.$ The
iso-spectrality condition gives rise to the AKNS hierarchy of nonlinear
dynamical systems on the functional manifold $M$ in the general form 
\begin{equation}
\frac{d}{dt_{j}}(u(t),v(t))^{\intercal }=K_{j}[u(t),v(t)],  \label{V1.2}
\end{equation}%
where 
\begin{equation}
\binom{u(t)}{v(t)}:=\binom{u(x+t_{1},t_{2,}t_{3,}...)}{%
v(x+t_{1},t_{2,}t_{3,}...)}  \label{V1.2a}
\end{equation}%
for $t\in \mathbb{C}^{\mathbb{N}}.$

The corresponding vector fields $K_{j}:M\rightarrow T(M),j\in \mathbb{N},$
can be constructed \cite{Bl,FT,HPP,Ne,PM,RS-T} via the following
Lie-algebraic scheme.

We define the centrally extended affine current $\mathfrak{s\ell }%
(2)-algebra\ \hat{\mathcal{G}}:=\tilde{\mathcal{G}}\oplus \mathbb{C}$ 
\begin{equation}
\tilde{\mathcal{G}}:=\{a=\sum_{j\in \mathbb{Z},\,j\ll \infty }a^{(j)}\otimes
\lambda ^{j}:a^{(j)}\in C^{\infty }\left( \mathbb{R}/2\pi \mathbb{Z};%
\mathfrak{s\ell }(2;\mathbb{C})\right) \mathbb{\}},  \label{V1.3}
\end{equation}%
endowed with the Lie commutator 
\begin{equation}
\lbrack (a_{1},c_{1}),(a_{2},c_{2})]:=([a_{1},a_{2}],\left\langle
a_{1},da_{2}/dx\right\rangle )  \label{V1.4}
\end{equation}%
with the scalar product 
\begin{equation}
\left\langle a_{1},a_{2}\right\rangle :=\mathrm{res}_{\lambda =\infty
}\int_{0}^{2\pi }\mathrm{tr}(a_{1}a_{2})dx  \label{V1.4a}
\end{equation}%
for any two elements $a_{1},a_{2}\in \tilde{\mathcal{G}}$, where "$\mathrm{%
res"}$ and "$\mathrm{tr"}$ are the usual residue and trace maps,
respectively. As the spectrum $\sigma (\ell )\subset \mathbb{C}$ is supposed
to be parametrically independent, there is a natural association with flows.
These flows are generated by the set $I(\hat{\mathcal{G}^{\ast }})$ of
Casimir invariants of the coadjoint action of the current algebra \ $\hat{%
\mathcal{G}}$ \ on a given element $\ell (x;\lambda )\in \tilde{\mathcal{G}}%
_{-}^{\ast }\cong \tilde{\mathcal{G}}_{+}$ contained in the space of
functionals $\mathcal{D}(\hat{\mathcal{G}})$. Here we have denoted by $%
\tilde{\mathcal{G}}:=\tilde{\mathcal{G}}_{+}\oplus \tilde{\mathcal{G}}_{-}$
the natural splitting into two affine subalgebras of positive and negative $%
\lambda $-expansions. In particular, a functional $\gamma (\lambda )\in I(%
\hat{\mathcal{G}})$ if and only if 
\begin{equation}
\lbrack \tilde{S}(x;\lambda ),\ell (x;\lambda )]+\frac{d}{dx}\tilde{S}%
(x;\lambda )=0,  \label{V1.5}
\end{equation}%
where the gradient $\tilde{S}(x;\lambda ):=\mathrm{grad}\gamma (\lambda
)(\ell )\in \tilde{\mathcal{G}}_{-}$ is defined with respect to the scalar
product (\ref{V1.4a}) by means of the variation 
\begin{equation}
\delta \gamma (\lambda ):=\left\langle \mathrm{grad}\gamma (\lambda )(\ell
),\delta \ell \right\rangle .  \label{V1.6}
\end{equation}%
We note here that the determining matrix equation (\ref{V1.5}) in the case
of the element $\ell (x;\lambda )\in \tilde{\mathcal{G}}_{-}^{\ast },$ given
by the spectral problem (\ref{V1.1}), can be easily solved recursively as $%
\lambda \rightarrow \infty $ in the following asymptotic form as%
\begin{align}
\tilde{S}(x;\lambda )& \sim \sum_{j\in \mathbb{Z}_{+}}\tilde{S}%
^{(j)}(x)\lambda ^{-(j+1)},\text{ \ \ }\tilde{S}(x;\lambda )=\left( 
\begin{array}{cc}
\tilde{S}_{11} & \tilde{S}_{12} \\ 
\tilde{S}_{21} & \tilde{S}_{22}%
\end{array}%
\right) ,  \label{V1.7} \\
\tilde{S}^{(0)}(x)& =\left( 
\begin{array}{cc}
1/2 & 0 \\ 
0 & -1/2%
\end{array}%
\right) ,\text{ \ }\tilde{S}^{(1)}(x)=\left( 
\begin{array}{cc}
0 & u \\ 
v & 0%
\end{array}%
\right) ,  \notag \\
\tilde{S}^{(2)}(x)& =\left( 
\begin{array}{cc}
-uv & u_{x} \\ 
-v_{x} & vu%
\end{array}%
\right) ,\text{ \ }\tilde{S}^{(3)}(x)=\left( 
\begin{array}{cc}
vu_{x}-uv_{x} & u_{xx}-2u^{2}v \\ 
v_{xx}-2v^{2}u & uv_{x}-vu_{x}%
\end{array}%
\right) ,...,  \notag
\end{align}%
and so on, based upon the differential relationships%
\begin{align}
\lambda \tilde{S}_{12}& =\tilde{S}_{12,x}+u(\tilde{S}_{11}-\tilde{S}_{22}), 
\notag \\
-\lambda \tilde{S}_{21}& =\tilde{S}_{21,x}-v(\tilde{S}_{11}-\tilde{S}_{22}),
\label{V1.8} \\
\tilde{S}_{11,x}& =u\tilde{S}_{21}-v\tilde{S}_{12}=-\tilde{S}_{22,x,}  \notag
\end{align}%
following from (\ref{V1.5}).

Now we will take into account that the coadjoint orbits of elements $\ell
\in \tilde{\mathcal{G}}_{-}^{\ast }$ with respect to the standard $\mathcal{R%
}$-structure \cite{FT} on the Lie\ algebra \ $\hat{\mathcal{G}}$ \ 
\begin{equation}
\lbrack (a_{1},c_{1}),(a_{2},c_{2})]_{\mathcal{R}}:=([\mathcal{R}%
a_{1},a_{2}]+[a_{1},\mathcal{R}a_{2}],\left\langle \mathcal{R}%
a_{1},da_{2}/dx\right\rangle -<da_{1}/dx,\mathcal{R}a_{2}>)  \label{V1.8a}
\end{equation}%
where, by definition, $\mathcal{R}:=\frac{1}{2}(P_{+}-P_{-})$ and $P_{\pm }%
\tilde{\mathcal{G}}:=\tilde{\mathcal{G}}_{\pm },$ are Poissonian manifolds 
\cite{AM,Ar,Bl,FT,HK,PM,RS-T}.\ Then the corresponding \textit{a priori}
iso-spectral AKNS flows can be constructed as the commuting Hamiltonian
systems on $\tilde{\mathcal{G}}_{-}^{\ast }$ $\ $\ 
\begin{equation}
\frac{d\ell }{dt_{j}}:=\{\gamma _{j},\ell \}=[(\lambda ^{j+1}\tilde{S}%
)_{+},\ell ]+\frac{d}{dx}(\lambda ^{j+1}\tilde{S})_{+}  \label{V1.9}
\end{equation}%
generated by the Casimir invariants $\gamma _{j}\in I(\hat{\mathcal{G}^{\ast
}}),j\in \mathbb{N},$ \ with respect to the Lie-Poisson structure on $\hat{%
\mathcal{G}}^{\ast }$ defined as%
\begin{equation}
\begin{array}{c}
\{\gamma ,\xi \}:=\left\langle \ell ,[\mathrm{grad}\gamma (\ell ),\mathrm{%
grad}\xi (\ell )]_{\mathcal{R}}\right\rangle + \\ 
+\left\langle \mathcal{R}\mathrm{grad}\gamma (\ell ),\frac{d}{dx}\mathrm{grad%
}\xi (\ell )\right\rangle -\left\langle \frac{d}{dx}\mathrm{grad}\gamma
(\ell ),\mathcal{R}\mathrm{grad}\xi (\ell )\right\rangle%
\end{array}
\label{V1.9a}
\end{equation}%
for any smooth functionals $\gamma ,\xi \in \mathcal{D}(\hat{\mathcal{G}}%
^{\ast }).$ \ As a result of (\ref{V1.9}) equation (\ref{V1.5}) is easily
augmented by the commuting hierarchy of evolution equations 
\begin{equation}
d\tilde{S}/dt_{j}=[(\lambda ^{j+1}\tilde{S})_{+},\tilde{S}]  \label{V1.10}
\end{equation}%
for $j\in \mathbb{N},$ including the determining equation (\ref{V1.5}) at $%
j=1.$

The hierarchy (\ref{V1.10}) can be rewritten with respect to the unique $%
\lambda $-parametric vector field 
\begin{equation}
d/dt:=\sum_{j\in \mathbb{Z}_{+}}\lambda ^{-j}d/dt_{j+1}  \label{V1.11}
\end{equation}%
on the manifold $M$ as the generating flow on $\mathcal{\tilde{G}}_{-}^{\ast
}:$%
\begin{equation}
\frac{d}{dt}\tilde{S}(x;\mu )=[\tilde{S}(x;\mu ),\frac{\lambda ^{3}}{\mu
-\lambda }\tilde{S}(x;\lambda )+\lambda \tilde{S}_{0}(x)],  \label{V1.12}
\end{equation}%
where $\mathbb{Z}_{+}:=\{{0\cup \mathbb{N\}}},$ the parameters $\lambda ,\mu
\rightarrow \infty $ in such a way that $|\mu /\lambda |<1.$ Since the flow (%
\ref{V1.9}) is, by construction, Hamiltonian on the adjoint space $\tilde{%
\mathcal{G}}_{-}^{\ast },$ it can be represented also as a Hamiltonian flow
on the functional manifold $M$ in the form \ 
\begin{equation}
\frac{d}{dt}\left( 
\begin{array}{c}
u \\ 
v%
\end{array}%
\right) =\left( 
\begin{array}{c}
\lambda ^{2}\tilde{S}_{12,x}+u\lambda ^{2}(\tilde{S}_{11}-\tilde{S}_{22}) \\ 
\lambda ^{2}\tilde{S}_{21,x}-v\lambda ^{2}(\tilde{S}_{11}-\tilde{S}_{22})%
\end{array}%
\right) ,  \label{V1.13}
\end{equation}%
which will be important for our further analysis. The representation \ (\ref%
{V1.13}) will be derived in the next two section with respect to both the
evolution vector field (\ref{V1.11}) and the related vertex vector field
mapping $X_{\lambda }:M\rightarrow M$ \ defined as 
\begin{align}
X_{\lambda }& :=(X_{\lambda }^{+},X_{\lambda }^{-}),\text{ \ \ }X_{\lambda
}^{+}=\exp D_{\lambda },\text{ \ \ }X_{\lambda }^{-}=\exp (-D_{\lambda }),%
\text{ \ \ \ \ \ }  \notag \\
\text{\ }D_{\lambda }& :=\sum_{j\in \mathbb{Z}_{+}}\frac{1}{(j+1)}\lambda
^{-(j+1)}\frac{d}{dt_{j+1}},  \label{V1.14}
\end{align}%
and satisfying the determining relationship 
\begin{equation}
\frac{d}{dt}=\mp \lambda ^{2}X_{\lambda }^{\pm ,-1}\frac{d}{d\lambda }%
X_{\lambda }^{\pm },  \label{V1.15}
\end{equation}%
as $\lambda \rightarrow \infty .$ These vertex vector field maps and their
connections with integrability theory have been studied extensively by a
number of researchers, most notably in \cite{Di,Ne}.

\section{Hamiltonian analysis}

Consider the Casimir functional $\gamma (\lambda )\in I(\hat{\mathcal{G}}),$ 
$\lambda \in \mathbb{C},$ and its gradient with respect to its dependence on
a point $(u,v)^{\intercal }\in M$ given by%
\begin{equation}
\mathrm{grad}\gamma (\lambda )[u,v]=(\tilde{S}_{21}(x;\lambda ),\tilde{S}%
_{12}(x;\lambda ))^{\intercal }\in T^{\ast }(M),  \label{V2.1}
\end{equation}%
as follows easily from definition (\ref{V1.6}). By introducing on the
manifold $M$ the following two skew-symmetric operators 
\begin{equation}
\theta :=\left( 
\begin{array}{cc}
0 & 1 \\ 
-1 & 0%
\end{array}%
\right) ,\text{ \ }\eta :=\left( 
\begin{array}{cc}
2u\partial ^{-1}u & \partial -2u\partial ^{-1}v \\ 
\partial -2v\partial ^{-1}u & 2v\partial ^{-1}v%
\end{array}%
\right) ,  \label{V2.2}
\end{equation}%
following easily \cite{FT,PM,PSAMP} from \ (\ref{V1.9a}), the relationships (%
\ref{V1.8}) can be rewritten as 
\begin{equation}
\lambda \theta \mathrm{grad}\gamma (\lambda )[u,v]=\eta \mathrm{grad}\gamma
(\lambda )[u,v],  \label{V2.3}
\end{equation}%
holding for all $\lambda \in \mathbb{C}.$ \ It is easy to verify that owing
to \ (\ref{V2.3}) the Casimir invariant $\gamma (\lambda )\in I(\hat{%
\mathcal{G}})$ simultaneously satisfies the two involutivity conditions%
\begin{equation}
\{\gamma (\lambda ),\gamma (\mu )\}_{\theta }=0=\{\gamma (\lambda ),\gamma
(\mu )\}_{\eta }  \label{V2.3a}
\end{equation}%
for all $\lambda ,\mu \in \mathbb{C}$ with respect to two Poissonian
structures 
\begin{equation}
\{\cdot ,\cdot \}_{\theta }:=(\mathrm{grad}(\cdot ),\theta \mathrm{grad}%
(\cdot )),\text{ \ \ }\{\cdot ,\cdot \}_{\eta }:=(\mathrm{grad}(\cdot ),\eta 
\mathrm{grad}(\cdot ))  \label{V2.6}
\end{equation}%
on the manifold $M,$ where $(\cdot ,\cdot )$ is the standard convolution on
the product bundle $T^{\ast }(M)\times T(M).$ \ \ As a direct consequence of
(\ref{V2.3}) and (\ref{V2.5}), one can readily verify that the operators $%
\theta ,\eta :T^{\ast }(M)\rightarrow T(M),$ defined by (\ref{V2.2}), are
co-symplectic, N\"{o}therian and compatible \cite{Bl,HPP,PM} on $M.$ This,
in particular, implies that the Lie derivatives \cite{AM,Ar,Bl,PM} 
\begin{equation}
L_{\frac{d}{dt}}\theta =0=L_{\frac{d}{dt}}\eta ,\text{ \ \ }L_{\frac{d}{dt}}%
\mathrm{grad}\gamma (\lambda )[u,v]=0  \label{V2.7}
\end{equation}%
vanish identically on the manifold $M.$\ 

Taking now into account (\ref{V1.9}) and \ (\ref{V2.2}), one finds easily
that%
\begin{equation}
\frac{d}{dt}\binom{u}{v}=-\{\lambda ^{2}\gamma (\lambda ),\binom{u}{v}%
\}_{\eta }=\lambda ^{2}\eta \mathrm{grad}\gamma (\lambda )[u,v]=\binom{%
\lambda ^{2}\tilde{S}_{12,x}+u\lambda ^{2}(\tilde{S}_{11}-\tilde{S}_{22})}{%
\lambda ^{2}\tilde{S}_{21,x}-v\lambda ^{2}(\tilde{S}_{11}-\tilde{S}_{22})}
\label{V2.4}
\end{equation}%
asymptotically as $\lambda \rightarrow \infty ,$ proving the representation
\ (\ref{V1.13}), mentioned in Introduction. Making use of the expansion \ (%
\ref{V1.7}) one easily obtains from \ (\ref{V2.4}) the first flows of the
AKNS hierarchy:%
\begin{equation}
\frac{d}{dt_{1}}\binom{u}{v}=\binom{u_{x}}{v_{x}}=K_{1}[u,v],\frac{d}{dt_{2}}%
\binom{u}{v}=\binom{u_{xx}+2u^{2}v}{-v_{xx}+2v^{2}u}=K_{2}[u,v],...,
\label{V2.5}
\end{equation}%
and so on. \ Below we will construct this infinite hierarchy of AKNS flows
by means of a very effective completely algebraical approach based on the
vertex operator representation of the solution to the generating flow \ (\ref%
{V2.4}).

\section{Vertex operator structure analysis}

It is well known \cite{FT,HPP,No,PM} that the Casimir invariants determining
equation (\ref{V1.5}) allows a general solution representation in the
following two important forms:%
\begin{equation}
\tilde{S}(x;\lambda )=k(\lambda )S(x;\lambda )-\frac{k(\lambda )}{2}\mathrm{%
tr}S(x;\lambda )  \label{V3.1}
\end{equation}%
and 
\begin{equation}
\tilde{S}(x;\lambda )=\tilde{F}(x,x_{0};\lambda )\tilde{C}(x_{0};\lambda )%
\tilde{F}^{-1}(x,x_{0};\lambda ).  \label{V3.1a}
\end{equation}%
Here, by definition, $S(x;\lambda ):=F(x+2\pi ,x;\lambda ),$ $F(y,x;\lambda
) $ and $\tilde{F}(y,x;\lambda )$ belong to the space of linear
endomorphisms of $\mathbb{C}^{2},$ $End$\thinspace $\mathbb{C}^{2},$ for all 
$x,x_{0},y\in \mathbb{R},$ and are matrix solutions to the spectral equation
(\ref{V1.1}) satisfying, respectively, the Cauchy problems%
\begin{equation}
\frac{\partial }{\partial y}F(y,x;\lambda )=\ell (y;\lambda )F(y,x;\lambda ),%
\text{ \ \ }F(y,x;\lambda )|_{y=x}=\mathbf{I,}  \label{V3.2}
\end{equation}%
and 
\begin{equation}
\frac{\partial }{\partial y}\tilde{F}(y,x;\lambda )=\ell (y;\lambda )\tilde{F%
}(y,x;\lambda ),\text{ \ \ }\tilde{F}(y,x;\lambda )|_{y=x}=\mathbf{I}%
+O(1/\lambda )\mathbf{,}  \label{V3.2a}
\end{equation}%
for all $\lambda \in \mathbb{C}\ $and $x\in \mathbb{R},$ where $\mathbf{I\in 
}$ $End$ $\mathbb{C}^{2}$ is the identity matrix. Here the parameters $%
k(\lambda )\in \mathbb{C}$ and $\tilde{C}(x_{0};\lambda )\in End$\thinspace $%
\mathbb{C}^{2}$ are invariant with respect to the generating vector field \ (%
\ref{V1.11}), chosen in such a way that the asymptotic condition 
\begin{equation}
\tilde{S}(x;\lambda )\in \tilde{\mathcal{G}}_{-}  \label{V3.3}
\end{equation}%
as $\lambda \rightarrow \infty $ holds for all $x\in \mathbb{R}.$

To construct the solution (\ref{V3.1}) satisfying condition (\ref{V3.3}), we
find a preliminary partial solution $\tilde{F}(y,x;\lambda )\in End$%
\thinspace $\mathbb{C}^{2},$ $x,y\in \mathbb{R},$ to equation (\ref{V3.2a})
\ satisfying the asymptotic Cauchy data 
\begin{equation}
\tilde{F}(y,x;\lambda )|_{y=x}=\mathbf{I}+O(1/\lambda )  \label{V3.4}
\end{equation}%
as $\lambda \rightarrow \infty .$ It is easy to check that 
\begin{equation}
\tilde{F}(y,x;\lambda )=\left( 
\begin{array}{cc}
\tilde{e}_{1}(y,x;\lambda ) & -\tilde{u}(y;\lambda )\lambda ^{-1}\tilde{e}%
_{2}(y,x;\lambda ) \\ 
\tilde{v}(y;\lambda )\lambda ^{-1}\tilde{e}_{1}(y,x;\lambda ) & \tilde{e}%
_{2}(y,x;\lambda )%
\end{array}%
\right)  \label{V3.5}
\end{equation}%
is an exact functional solution to (\ref{V3.2a}) satisfying condition (\ref%
{V3.4}). Here we have defined 
\begin{align}
\tilde{e}_{1}(y,x;\lambda )& :=\exp \{(y-x)\lambda /2+\lambda
^{-1}\int_{x}^{y}u\tilde{v}ds\},  \label{V3.6} \\
\tilde{e}_{2}(y,x;\lambda )& :=\exp \{(x-y)\lambda /2-\lambda
^{-1}\int_{x}^{y}\tilde{u}vds\},  \notag
\end{align}%
where the vector-function $(\tilde{u},\tilde{v})^{\intercal }\in M$ \
satisfies the determining functional relationships%
\begin{equation}
\tilde{u}=u+\tilde{u}_{x}\lambda ^{-1}-\tilde{u}^{2}v\lambda ^{-2},\text{ \
\ \ }\tilde{v}=v-\tilde{v}_{x}\lambda ^{-1}-\tilde{v}^{2}u\lambda ^{-2},
\label{V3.7}
\end{equation}%
as $\lambda \rightarrow \infty ,$ which were discovered earlier in a very
interesting article \cite{PV}. There was also shown that exact asymptotic
(as $\lambda \rightarrow \infty $) functional solutions of these
relationships can be easily constructed by means of the standard iteration
procedure.

The fundamental matrix $F(y,x;\lambda )\in End$\thinspace $\mathbb{C}^{2}$
is represented for all $x,y\in \mathbb{R}$ in the form%
\begin{equation}
F(y,x;\lambda )=\tilde{F}(y,x;\lambda )\tilde{F}^{-1}(x,x;\lambda ).
\label{V3.8}
\end{equation}%
Consequently, if one sets $y=x+2\pi $ in this formula and defines 
\begin{equation}
k(\lambda ):=\lambda ^{-1}[\tilde{e}_{1}(x+2\pi ,x;\lambda )-\tilde{e}%
_{2}(x+2\pi ,x;\lambda )]^{-1},  \label{V3.8a}
\end{equation}%
it follows from (\ref{V3.8}) that the exact matrix representation%
\begin{equation}
\tilde{S}(x;\lambda )=\left( 
\begin{array}{cc}
\frac{\lambda ^{2}-\tilde{u}\tilde{v}}{2\lambda (\lambda ^{2}+\tilde{u}%
\tilde{v})} & \frac{\tilde{u}}{\lambda ^{2}+\tilde{u}\tilde{v}} \\ 
\frac{\tilde{v}}{\lambda ^{2}+\tilde{u}\tilde{v}} & \frac{\tilde{u}\tilde{v}%
-\lambda ^{2}}{2\lambda (\lambda ^{2}+\tilde{u}\tilde{v})}%
\end{array}%
\right)  \label{V3.9}
\end{equation}%
satisfies the necessary condition (\ref{V3.3}) as $\lambda \rightarrow
\infty $ .

\begin{remark}
\label{Rm_V1}The invariance of the functional (\ref{V3.8a}) with respect to
the generating vector field (\ref{V1.11}) on the manifold $M$ derives from
the representation (\ref{V3.5}), the evolution equations (\ref{V3.2a}) and 
\begin{equation}
\frac{d}{dt}\tilde{F}(y,x;\mu )=\left( \frac{\lambda ^{3}}{\mu -\lambda }%
\tilde{S}(x;\lambda )+\lambda \tilde{S}_{0}(x)\right) \tilde{F}(y,x;\mu ),
\label{V3.10}
\end{equation}%
which follows naturally from the determining matrix flows (\ref{V1.9}) upon
applying the translation $y\rightarrow y+2\pi .$
\end{remark}

The matrix expression (\ref{V3.9}) coincides as $\lambda \rightarrow \infty $
with the asymptotic expansion (\ref{V1.7}), whose matrix elements satisfy
the following important functional relationships:%
\begin{equation}
\frac{1-\lambda (\tilde{S}_{11}-\tilde{S}_{22})}{2\tilde{S}_{21}}=\tilde{u},%
\text{ }\frac{1-\lambda (\tilde{S}_{11}-\tilde{S}_{22})}{2\tilde{S}_{12}}=%
\tilde{v},  \label{V3.11}
\end{equation}%
allowing the introduction in a natural way of the vertex vector field (\ref%
{V1.14}). To show this, we need to take the preliminary step of deriving the
corresponding evolution equation for the vector function $(\tilde{u},\tilde{v%
})^{\intercal }\in M$ \ with respect to the generating vector field (\ref%
{V1.11}) in the asymptotic form (\ref{V1.12}) as $\lambda \rightarrow \infty
.$ Before doing this we shall find the form of evolution equation (\ref%
{V1.13}) as $\mu ,\lambda \rightarrow \infty :$%
\begin{equation}
\frac{d}{dt}\tilde{S}(x;\mu )=[\lambda ^{3}\frac{d}{d\lambda }\tilde{S}%
(x;\mu )-\lambda \tilde{S}_{0}(x),\tilde{S}(x;\lambda )],  \label{V3.12}
\end{equation}%
which entails the following differential relationships: 
\begin{equation}
\begin{array}{c}
d\tilde{S}_{11}/dt=\lambda ^{3}(\tilde{S}_{21}d\tilde{S}_{12}/d\lambda -%
\tilde{S}_{12}d\tilde{S}_{21}/d\lambda ), \\ 
d\tilde{S}_{22}/dt=\lambda ^{3}(\tilde{S}_{12}d\tilde{S}_{21}/d\lambda -%
\tilde{S}_{21}d\tilde{S}_{12}/d\lambda ), \\ 
d\tilde{S}_{12}/dt=\lambda ^{3}[\tilde{S}_{12}\frac{d}{d\lambda }(\tilde{S}%
_{11}-\tilde{S}_{22})-(\tilde{S}_{11}-\tilde{S}_{22})\frac{d\tilde{S}_{12}}{%
d\lambda })-\lambda \tilde{S}_{12}, \\ 
d\tilde{S}_{21}/dt=\lambda ^{3}[\tilde{S}_{21}\frac{d}{d\lambda }(\tilde{S}%
_{22}-\tilde{S}_{11})-(\tilde{S}_{22}-\tilde{S}_{11})\frac{d\tilde{S}_{21}}{%
d\lambda })+\lambda \tilde{S}_{21}.%
\end{array}
\label{V3.13}
\end{equation}%
Using the relationships (\ref{V3.13}), one can easily obtain by means of
simple, but rather cumbersome calculations, the evolution equations for the
vector function $(\tilde{u},\tilde{v})^{\intercal }\in M$ \ expressed in the
form (\ref{V3.11})%
\begin{equation}
\begin{array}{c}
\frac{d}{dt}[\frac{1-\lambda (\tilde{S}_{11}-\tilde{S}_{22})}{2\tilde{S}_{12}%
}]=-\lambda ^{2}\frac{d}{d\lambda }[\frac{1-\lambda (\tilde{S}_{11}-\tilde{S}%
_{22})}{2\tilde{S}_{12}}],\text{ } \\ 
\frac{d}{dt}[\frac{1-\lambda (\tilde{S}_{11}-\tilde{S}_{22})}{2\tilde{S}_{12}%
}]=\lambda ^{2}\frac{d}{d\lambda }[\frac{1-\lambda (\tilde{S}_{11}-\tilde{S}%
_{22})}{2\tilde{S}_{12}}],%
\end{array}
\label{V3.14}
\end{equation}%
which hold as $\lambda \rightarrow \infty .$ As a direct consequence of the
differential relationships (\ref{V3.14}), the following \textit{vertex
operator representation }for the vector function $(\tilde{u},\tilde{v}%
)^{\intercal }\in M$%
\begin{align}
\tilde{u}(t;\lambda )& :=u^{+}(t;\lambda )=X_{\lambda }^{+}u(t),
\label{V3.15} \\
\tilde{v}(t;\lambda )& :=v^{-}(t;\lambda )=X_{\lambda }^{-}u(t),  \notag
\end{align}%
holds. Here we took into account that, owing to the determining functional
representations (\ref{V3.7}), the limits 
\begin{equation}
\lim_{\lambda \rightarrow \infty }\tilde{u}(t;\lambda )=u(t),\text{ \ \ }%
\lim_{\lambda \rightarrow \infty }\tilde{v}(t;\lambda )=v(t),  \label{V3.16}
\end{equation}%
exist and the vertex operator $\ X_{\lambda }:M\rightarrow M$ acts on the
functional manifold $M$ via the corresponding shift operators defined above
by means of the differential relationships (\ref{V1.14}) and (\ref{V1.15}).
Moreover, from \ (\ref{V3.7}) one obtains that 
\begin{equation}
u^{+}=u+u_{x}^{+}\lambda ^{-1}-(u^{+})^{2}v\lambda ^{-2},\ \ \
v^{-}=v-v_{x}^{-}\lambda ^{-1}-(v^{-})^{2}u\lambda ^{-2},  \label{V3.16a}
\end{equation}

The vertex representation (\ref{V3.16a}) allows, in particular, to readily
construct infinite hierarchies of the conservation laws for the generating
AKNS integrable vector field (\ref{V1.11}) as 
\begin{equation}
H_{+}(\lambda ):=\int_{0}^{2\pi }u^{+}(t;\lambda )v(t)dx,\text{ }%
H_{-}(\lambda ):=\int_{0}^{2\pi }v^{-}(t;\lambda )u(t)dx,  \label{V3.17}
\end{equation}%
which follow from (\ref{V3.5}), (\ref{V3.6}) and reasoning from Remark (\ref%
{Rm_V1}). Since the fundamental matrix (\ref{V3.8}) at $y=x+2\pi $ defines
via relationship \ (\ref{V3.1}) the solution 
\begin{equation}
S(x;\lambda ):=\tilde{F}(x+2\pi ,x;\lambda )\tilde{F}^{-1}(x,x;\lambda )
\label{V3.18}
\end{equation}%
to the determining equations (\ref{V1.5}) and (\ref{V1.8}), its determinant $%
\det S(x;\lambda )$ is invariant with respect to the generating vector field
(\ref{V1.11}) and equals $\det S(x;\lambda )=\det \tilde{F}(x+2\pi
,x;\lambda )\det \tilde{F}^{-1}(x,x;\lambda )=1$ for all $x\in \mathbb{R}$
and $\lambda \in \mathbb{C}$ owing to the condition $\mathrm{tr}$ $\ell
(x;\lambda )=0.$ Accordingly, based on the matrix representation (\ref{V3.5}%
), one finds that the relationships 
\begin{equation}
\begin{array}{c}
\tilde{e}_{1}(x+2\pi ,x;\lambda ):=\exp \left[ \pi \lambda +\lambda
^{-1}H_{+}(\lambda )\right] , \\ 
\tilde{e}_{2}(x+2\pi ,x;\lambda ):=\exp \left[ -\pi \lambda -\lambda
^{-1}H_{-}(\lambda )\right] , \\ 
\tilde{e}_{1}(x+2\pi ,x;\lambda )\tilde{e}_{2}(x+2\pi ,x;\lambda )=1, \\ 
\frac{d}{dt}\tilde{e}_{1}(x+2\pi ,x;\lambda )=0=\frac{d}{dt}\tilde{e}%
_{2}(x+2\pi ,x;\lambda )%
\end{array}
\label{V3.19}
\end{equation}%
hold for all $x\in \mathbb{R}$ and $\lambda \in \mathbb{C}.$ As a
consequence of (\ref{V3.19}), we obtain 
\begin{equation}
H_{+}(\lambda )=H_{-}(\lambda )  \label{V3.20}
\end{equation}%
for all $\lambda \in \mathbb{C}$; that is, the two hierarchies of
conservations law (\ref{V3.17}) coincide. Concerning the AKNS hierarchy
vector fields (\ref{V1.11}) and the related Hamiltonian flows on the
manifold $M,$ we can easily derive them from the canonical vertex
representations (\ref{V3.15}), taking into account the recursive functional
equations (\ref{V3.7}). We obtain from that (\ref{V3.7}) and (\ref{V3.17})
that%
\begin{eqnarray}
X_{\lambda }^{+}u &=&u^{+}=u+\lambda ^{-1}u_{x}+\lambda
^{-2}[u_{xx}^{+}+(u^{+})^{2}v]+\lambda ^{-3}[(u^{+})^{2}v]_{x}=...,
\label{V3.21} \\
X_{\lambda }^{-}v &=&v^{-}=v-\lambda ^{-1}v_{x}-\lambda
^{-2}[v_{xx}^{-}+(v^{-})^{2}u]+\lambda ^{-3}[(v^{-})^{2}u]_{x}=...,  \notag
\end{eqnarray}%
which immediately yield the whole AKNS hierarchy of nonlinear Lax integrable
dynamical systems on the functional manifold $M.$ For instance, we obtain
from \ (\ref{V3.21}) the AKNS flows 
\begin{equation}
\frac{d}{dt_{1}}\binom{u}{v}=\binom{u_{x}}{v_{x}},\frac{d}{dt_{2}}\binom{u}{v%
}=\binom{u_{xx}+2u^{2}v}{-v_{xx}+2v^{2}u},...,  \label{V3.21a}
\end{equation}%
and so on, which coincide with those constructed before in \ (\ref{V2.5}).

\section{The $\protect\tau $-function representation}

The vertex operator representations (\ref{V3.5}), (\ref{V3.9}) and (\ref%
{V3.15}) can also be naturally associated with the results in \cite{Di,Ne},
based on the generating $\tau $-function approach. The latter makes
extensive use of the versatile the dual representation (\ref{V3.1a}) for the
generating current algebra element $\tilde{S}(x;\lambda )\in \tilde{\mathcal{%
G}}_{-}^{\ast }$ $\ ($as $\lambda \rightarrow \infty )$ for the AKNS flows
with the specially chosen invariant matrix 
\begin{equation}
\tilde{C}(x_{0};\lambda )=\left( 
\begin{array}{cc}
1 & 0 \\ 
0 & -1%
\end{array}%
\right) \in End\,\mathbb{C}^{2}.  \label{V3.21b}
\end{equation}%
In the context of our approach, the relation with the $\tau $-function
representation devised in \cite{Di,Ne} can be based on the matrix solution (%
\ref{V3.5}) and the simple vertex operator mapping properties%
\begin{equation}
X_{\lambda }\left( 
\begin{array}{c}
\tilde{e}_{1}(x,y;\lambda ) \\ 
\tilde{e}_{2}(x,y;\lambda )%
\end{array}%
\right) =\left( 
\begin{array}{c}
\tilde{e}_{2}(y,x;\lambda ) \\ 
\tilde{e}_{1}(y,x;\lambda )%
\end{array}%
\right) ,\text{ }  \label{V3.22}
\end{equation}%
which follow directly from the definitions (\ref{V3.6}) and (\ref{V3.15}). \
Based on the functional relationships \ \ (\ref{V3.6}) and (\ref{V3.16a}),
one easily obtains the following differential expresions: 
\begin{equation}
\tilde{F}(y,x;\lambda )=\left( 
\begin{array}{cc}
\tilde{e}_{1}(y,x;\lambda ) & -u^{+}(y;\lambda )\lambda ^{-1}\tilde{e}%
_{2}(y,x;\lambda ) \\ 
v^{-}(y;\lambda )\lambda ^{-1}\tilde{e}_{1}(y,x;\lambda ) & \tilde{e}%
_{2}(y,x;\lambda )%
\end{array}%
\right)  \label{V3.22a}
\end{equation}%
and 
\begin{equation}
\lambda (u^{+}v^{+}-u^{-}v^{-})=\frac{\partial }{\partial y}(u^{+}v+uv^{-}),
\label{V3.22b}
\end{equation}%
where 
\begin{align}
\tilde{e}_{1}(y,x;\lambda )& :=\exp \{(y-x)\lambda /2+\lambda
^{-1}\int_{x}^{y}uv^{-}ds\},  \label{V3.22c} \\
\tilde{e}_{2}(y,x;\lambda )& :=\exp \{(x-y)\lambda /2-\lambda
^{-1}\int_{x}^{y}u^{+}vds\},  \notag
\end{align}%
Moreover, \ one can easily derive from \ (\ref{V3.22c}) that 
\begin{equation}
\lambda \frac{\partial ^{2}}{\partial y^{2}}\ln \frac{\tilde{e}%
_{1}(y,x;\lambda )}{\tilde{e}_{2}(y,x;\lambda )}=\frac{\partial }{\partial y}%
(u^{+}v+uv^{-}),  \label{V3.22d}
\end{equation}%
giving rise jointly with \ (\ref{V3.22}) the \ importnat functional
relationship 
\begin{equation}
\frac{\partial ^{2}}{\partial y^{2}}\ln \frac{\tilde{e}_{1}(y,x;\lambda )}{%
\tilde{e}_{2}(y,x;\lambda )}=(u^{+}v^{+}-u^{-}v^{-}),  \label{V3.22e}
\end{equation}%
which allows to introduce naturally a so-called $\tau $-function \
repreentation 
\begin{equation}
\frac{\tilde{e}_{1}(y,x;\lambda )}{\tilde{e}_{2}(y,x;\lambda )}:=\frac{\tau
^{-}(y,x;\lambda )}{\tau ^{+}(y,x;\lambda )}\exp [\alpha (x;\lambda )+y\beta
(x;\lambda )]  \label{V3.22f}
\end{equation}%
for some functions $\alpha (\cdot ;\lambda ),\beta (\cdot ;\lambda ):\mathbb{%
R\rightarrow C},$ \ where we put, by definition, 
\begin{equation}
-\frac{\partial ^{2}}{\partial y^{2}}\ln \tau :=uv,  \label{V3.22g}
\end{equation}%
which coincides with that, presented in \cite{Ne,MJD}. Taking now into
account the relationships \ (\ref{V3.22}), one easily obtains that 
\begin{eqnarray}
\tilde{e}_{1}(y,x;\lambda ) &=&\exp [\alpha (x;\lambda )+\frac{\lambda }{2}%
(y-x)]\tau ^{-}(y,x;\lambda )/\tau (y,x;\lambda ),  \label{V3.22h} \\
\tilde{e}_{2}(y,x;\lambda ) &=&\exp [\alpha (x;\lambda )+\frac{\lambda }{2}%
(x-y)]\tau ^{+}(y,x;\lambda )/\tau (y,x;\lambda )  \notag
\end{eqnarray}%
for all $x,y\in \mathbb{R}$ and $\lambda \in \mathbb{C}.$ \ As a result of (%
\ref{V3.5}) and (\ref{V3.22h}), the crucial expression for the normalized
matrix 
\begin{align}
\ \bar{F}(y,x;\lambda )& :=\left( \det \tilde{F}(x,x;\lambda )\right) ^{-1/2}%
\tilde{F}(y,x;\lambda )=  \notag \\
& =\left( 
\begin{array}{cc}
\frac{\lambda \tilde{e}_{2}^{-}(x,y;\lambda )}{[\lambda ^{2}+u^{+}(x;\lambda
)v^{-}(x;\lambda )]^{1/2}} & -\frac{u^{+}(y;\lambda )\tilde{e}%
_{1}^{+}(x,y;\lambda )}{[\lambda ^{2}+u^{+}(x;\lambda )v^{-}(x;\lambda
)]^{1/2}} \\ 
\frac{v^{-}(y;\lambda )\tilde{e}_{2}^{+}(x,y;\lambda )}{[\lambda
^{2}+u^{+}(x;\lambda )v^{-}(x;\lambda )]^{1/2}} & \frac{\lambda \tilde{e}%
_{1}^{+}(x,y;\lambda )}{[\lambda ^{2}+u^{+}(x;\lambda )v^{-}(x;\lambda
)]^{1/2}}%
\end{array}%
\right) :=  \label{V3.23} \\
& =\left( 
\begin{array}{cc}
\frac{\tau ^{-}(y,x;\lambda )}{\tau (y,x;\lambda )}\exp [\frac{\lambda }{2}%
(y-x)]\  & -\frac{u^{+}(y;\lambda )\tau ^{+}(y,x;\lambda )}{\lambda \tau
(y,x;\lambda )}\exp [\frac{\lambda }{2}(x-y)] \\ 
\frac{v^{-}(y;\lambda )\tau ^{-}(y,x;\lambda )}{\lambda \tau (y,x;\lambda )}%
\exp [\frac{\lambda }{2}(y-x)] & \frac{\tau ^{+}(y,x;\lambda )}{\tau
(y,x;\lambda )}\exp [\frac{\lambda }{2}(x-y)]%
\end{array}%
\right) ,  \notag
\end{align}%
owing to \ (\ref{V3.22h}) holds, where 
\begin{align}
\frac{\tau ^{-}(y,x;\lambda )}{\tau (y,x;\lambda )}\exp [\frac{\lambda }{2}%
(y-x)]& :=\frac{\lambda \tilde{e}_{2}^{-}(x,y;\lambda )}{[\lambda
^{2}+u^{+}(x;\lambda )v^{-}(x;\lambda )]^{1/2}},  \notag \\
\frac{\tau ^{+}(y,x;\lambda )}{\tau (y,x;\lambda )}\exp [\frac{\lambda }{2}%
(x-y)]& :=\frac{\lambda \tilde{e}_{1}^{+}(x,y;\lambda )}{[\lambda
^{2}+u^{+}(x;\lambda )v^{-}(x;\lambda )]^{1/2}},  \label{V3.24}
\end{align}%
jointly with the compatibility relationship%
\begin{equation}
\exp [-\alpha (x;\lambda )]:=[\lambda ^{2}+u^{+}(x;\lambda )v^{-}(x;\lambda
)]^{1/2}.  \notag
\end{equation}

The vertex operator expression (\ref{V3.23}), as is easily checked, can be
readily employed to derive the representation (\ref{V3.1a}), where the exact
result (\ref{V3.9}) entails the additional application of the useful \cite%
{Ne} vertex representation%
\begin{equation}
\bar{F}(y,x;\lambda )=\left( 
\begin{array}{cc}
\frac{\tau ^{-}(y,x;\lambda )}{\tau (y,x;\lambda )}\exp [\frac{\lambda }{2}%
(y-x)] & -\frac{\sigma ^{+}(y,x;\lambda )}{\lambda \tau (y,x;\lambda )}\exp [%
\frac{\lambda }{2}(x-y)] \\ 
\frac{\rho ^{-}(y,x;\lambda )}{\lambda \tau (y,x;\lambda )}\exp [\frac{%
\lambda }{2}(y-x)] & \frac{\tau ^{+}(y,x;\lambda )}{\tau (y,x;\lambda )}\exp
[\frac{\lambda }{2}(x-y)]%
\end{array}%
\right) ,  \label{V3.25}
\end{equation}%
which holds as $\lambda \rightarrow \infty $ if $\rho (y,x;\lambda
):=v(y)\tau (y,x;\lambda ),\sigma (y,x;\lambda ):=u(y)\tau (y,x;\lambda ),$ $%
\ x,y\in \mathbb{R},$ and $\ $\ mappings $\ \rho ^{-}$ and $\sigma ^{+}$ are
defined in the obvious fashion. In this regard, it should be noted that the
vertex operator representation (\ref{V3.25}) for the matrix (\ref{V3.23})
was obtained in \cite{Ne} as a special normalized solution to the
determining equation (\ref{V3.2a}). Taking into account these two dual
vertex representations of the AKNS hierarchy of integrable flows on the
functional manifold $M,$ one can see that the first one - presented in this
work - is both technically simpler and more effective in obtaining exact
descriptions of such important functional ingredients as conservation laws,
symplectic structures and related commuting vector fields.

\section{Concluding remarks}

The vertex operator functional representations of the matrix solutions (\ref%
{V3.5}) and (\ref{V3.9}) for the determining equations (\ref{V3.2}) and (\ref%
{V1.5}), respectively, as one can see from the above analysis, are
essentially derived from the generating AKNS hierarchy vector field (\ref%
{V1.11}) on the functional manifold $M$ and its intrinsic Lie-algebraic
structure (\ref{V1.9}). As the decisive property of the vertex operator
relationships (\ref{V3.15}) and (\ref{V3.17}) is fundamentally based on the
representations (\ref{V3.11}) and equations (\ref{V3.13}), they provide a
very straightforward and transparent explanation of many of
\textquotedblleft miraculous\textquotedblright\ calculations in \cite%
{Di,Ne,MJD}. Of course, the results for the AKNS hierarchy in these and in
earlier papers \cite{Ne,MJD,BtK,DKJM,DS,SW,Wil} were obtained in a
distinctly different manner making use of direct asymptotic power series
expansions of solutions to the determining matrix equations (\ref{V3.2}) and
(\ref{V3.5}) and studying their deep algebraic properties.

It should be noted that, in a certain sense, the effectiveness of our
approach to studying the vertex operator representation of the AKNS
hierarchy owes much to the important exact representation (\ref{V3.1}) for
the solution of the Casimir invariants determining equation (\ref{V1.5}),
strongly based on the well known monodromy matrix approach devised \ by S.
Novikov's in \cite{No}. This equation entails the extremely effective AKNS
hierarchy representation in the simple recursive form (\ref{V3.21}), which
explains several other very interesting results in the literature, such as
in \cite{PV,Ve}. On the other hand, the dual solution representation to (\ref%
{V1.5}) in the form (\ref{V3.1}), used extensively in \cite{Ne}, led
naturally to the introduction of the well-known $\tau $-function and made it
possible to present the whole AKNS hierarchy in terms of its suitable
partial derivatives. Nonetheless, both our vertex operator approach and the $%
\tau $-function method, as was briefly demonstrated above, are intimately
related to each other.

As an indication of possible future research, it should also be mentioned
that it would be interesting to apply the vertex operator approach devised
in this work to other linear spectral problems such as those related to the
generalized Riemann hydrodynamical systems and BSR systems studied recently
in \cite{BPS,BUR,GPPP,PAPP}.

\section{Acknowledgments}

D. Blackmore wishes to thank the National Science Foundation for support
from NSF Grant CMMI - 1029809 and his coauthor for enlisting him in the
efforts that produced this paper. A.K. Prykarpatsky cordially thanks Profs.
M.V. Pavlov and V. H. Samoylenko for very useful discussions of the results
obtained. The last but not least our thanks go to Referees for their
instrumental comments and remarks which helped to improve the exposition of
the results.

\end{document}